\documentclass[showpacs,preprintnumbers]{revtex4}

\usepackage[centertags]{amsmath}
\usepackage{amsfonts}
\usepackage{amssymb}
\usepackage{amsthm}
\usepackage{newlfont}

\newtheorem{thm}{Theorem}[section]
\newtheorem{lem}[thm]{Lemma}
\newtheorem{defn}[thm]{Definition}

\begin{document}

\title{Simple test for quantum channel capacity}
\author{Marcin L. Nowakowski$^{1}$ and Pawel Horodecki$^{1}$\footnote{Electronic address: pawel@mif.pg.gda.pl}}
\affiliation{$^{1}$Faculty of Applied Physics and Mathematics,
~Gdansk University of Technology, 80-952 Gdansk, Poland}



\begin{abstract}
Basing on states and channels isomorphism we point out that
semidefinite programming can be used as a quick test for nonzero
one-way quantum channel capacity. This can be achieved by search
of symmetric extensions of states isomorphic to a given quantum
channel. With this method we provide examples of quantum channels
that can lead to high entanglement transmission but still have
zero one-way capacity, in particular, regions of symmetric
extendibility for isotropic states in arbitrary dimensions are
presented. Further we derive {\it a new entanglement parameter}
based on (normalised) relative entropy distance to the set of
states that have symmetric extensions and show explicitly the
symmetric extension of isotropic states being the nearest to
singlets in the set of symmetrically extendible states. The suitable
regularisation of the parameter provides a new upper bound on
one-way distillable entanglement.
\end{abstract}

\maketitle

\section{Introduction}

Quantum channels \cite{huge} are very important notion of quantum
information theory \cite{Springer}. It has been  proven
\cite{huge} time that there is a connection between entanglement
distillation \cite{distillation} and quantum channels capacities.
No-cloning principle has been used to prove that for some region
quantum depolarising channel has zero capacity even if does not
destroy entanglement \cite{Bruss}.

Following seminal work \cite{Barnum0} and asymptotic analysis
\cite{channels} that predicted  limit formulas form conjectured
hashing inequality \cite{channels} recently the latter have been
proven \cite{Shor,Devetak,DV04}, in particular the proof of
hashing inequality has been provided \cite{DV04}. On the other
hand the connection between quantum channels and entanglement
distillation \cite{huge} has been developed \cite{channels,PH03}
leading in particular to strong nonadditivity effects in case of
more than one receiver \cite{DCH04}. On the other hand an
interesting technique based on approximate quantum cloning was
used in Ref. \cite{Bruss} to point out limit of depolarising qubit
channel. This approach has been further extended in an elegant way
to the case of Pauli channels \cite{Cerf} via assymetric cloning
machines.

In the present approach we shall use the above techniques,
exploiting also a general notion of symmetric extension of quantum
state that was used recently efficiently applied with help of
semidefinite programming to characterise quantum entanglement
\cite{D1} \cite{D2} and states that admit local hidden variables
models \cite{T1}.

To be more specific, in this paper we develop qualitative
equivalence between entanglement distillation and quantum channels
theory showing in particular that:

(i) semidefinite programming can serve as a simple and quick test
for nonzero one-way channel capacity via looking
 for symmetric extensions of the state $\varrho(\Lambda)$,

(ii) if normalised and regularised, the distance of given quantum
state above to the set of symmetrically extendible states provides
{\it a new entanglement parameter} that leads to upper bound on
one-way distillable entanglement of the state.

To show that SDP can lead to interesting results we provide the
family of quantum channels that allow for  quite high entanglement
transmission, however have one-way capacity zero due to existence
of symmetric extension of the corresponding quantum state. The
corresponding extensions are explicitly constructed.

\section{One-way distillable entanglement}

Following the idea \cite{Bruss} developing restriction on qubit
depolarising channel from approximate quantum cloning we shall
utilise general notion of symmetric extensions of quantum state
(see \cite{D1,D2,T1}) to provide a general rule and examples of
channels with zero one-way capacity. We show now that every state
$\rho_{AB}(\Lambda)$ which has a symmetric extension $\rho_{ABB'}$
has special featured \textit{$D_{\rightarrow}$} and
\textit{$Q_{\rightarrow}$} according to its quantum channel
implied by Jamiolkowski isomorphism. The following observation
that describes above reads:

{\textbf{Observation 1.}} If any bipartite state $\rho_{AB}$ has a
symmetric extension $\rho_{ABB'}$, so that
$\rho_{ABB'}=\rho_{AB'B}$, then for the one-way distillable
entanglement there holds: $D_{\rightarrow}(\varrho_{AB})$=0.

Proof of the above theorem is immediate and follows from quantum
entanglement monogamy (cf. \cite{Bruss,Cerf}). If Alice sends
classical information to Bob and they distill singlet in the
protocol then the state can not have symmetric extension since
Bob's colleague, say Brigitte (corresponding to index B') could
also receive the same message from Alice and finally share a
singlet  with Alice  too. But Alice's particle can not be
maximally entangled  with two different particles at the some time
(this is just the entanglement monogamy property). So a
symmetrically extendible state can not have one-way distillable
entanglement nonzero. Combining the above observation we get
immediately

{\textbf{Observation 2.}} A sufficient condition for one-way
quantum capacity of given quantum channel $\Lambda$ to vanish is
symmetric extendibility of the state $\varrho(\Lambda)$ isomorphic
to the channel.

As a special example of application of these observations we use
below bipartite state $\rho_{AB}$ that is extendible for
$F\leq\frac{1}{2}$, moreover, notice that in this range the state
may be quite strong entangled.
\begin{equation}
  \rho_{AB} = \begin{pmatrix}
  \frac{F}{3} & 0 & 0 & 0 & \frac{F}{3} & 0 & 0 & 0 & \frac{F}{3} \\
  0 & \frac{(1-F)}{3} & 0 & 0 & 0 & 0 & 0 & 0 & 0 \\
  0 & 0 & 0 & 0 & 0 & 0 & 0 & 0 & 0 \\
  0 & 0 & 0 & 0 & 0 & 0 & 0 & 0 & 0 \\
  \frac{F}{3} & 0 & 0 & 0 & \frac{F}{3} & 0 & 0 & 0 & \frac{F}{3} \\
  0 & 0 & 0 & 0 & 0 & 0 & 0 & 0 & 0 \\
  0 & 0 & 0 & 0 & 0 & 0 & \frac{(1-F)}{3} & 0 & 0 \\
  0 & 0 & 0 & 0 & 0 & 0 & 0 & \frac{(1-F)}{3} & 0 \\
  \frac{F}{3} & 0 & 0 & 0 & \frac{F}{3} & 0 & 0 & 0 & \frac{F}{3}
\end{pmatrix}
\end{equation}

Notice that filtering on Bob's side the state $\rho_{AB}$ and in
general any such a state does not change the extendibility, what
may be simply proved. Applying filtering with $W=diag \left[
  1,\; \frac{1}{\sqrt{F}},\; \frac{1}{\sqrt{2-F}}
\right]$ we get a state $\widetilde{\rho}_{AB}$ and maximally
mixed $\widetilde{\rho_{A}}$ on Alice's side
\[
\widetilde{\rho}_{AB} = \frac{W\otimes I \rho_{AB} W^{\dag}\otimes
I}{Tr\{W\otimes I \rho_{AB} W^{\dag}\otimes I\}},\;
\widetilde{\rho}_{A} = \frac{I}{3}
\]
\begin{equation}
  \widetilde{\rho}_{AB} = \begin{pmatrix}
  \frac{F}{3} & 0 & 0 & 0 & \frac{\sqrt{F}}{3} & 0 & 0 & 0 & \frac{F}{3\sqrt{2-F}} \\
  0 & \frac{1-F}{3} & 0 & 0 & 0 & 0 & 0 & 0 & 0 \\
  0 & 0 & 0 & 0 & 0 & 0 & 0 & 0 & 0 \\
  0 & 0 & 0 & 0 & 0 & 0 & 0 & 0 & 0 \\
  \frac{\sqrt{F}}{3} & 0 & 0 & 0 & \frac{1}{3} & 0 & 0 & 0 & \frac{\sqrt{F}}{3\sqrt{2-F}} \\
  0 & 0 & 0 & 0 & 0 & 0 & 0 & 0 & 0 \\
  0 & 0 & 0 & 0 & 0 & 0 & \frac{1-F}{3(2-F)} & 0 & 0 \\
  0 & 0 & 0 & 0 & 0 & 0 & 0 & \frac{1-F}{3(2-F)} & 0 \\
  \frac{F}{3\sqrt{2-F}} & 0 & 0 & 0 & \frac{\sqrt{F}}{3\sqrt{2-F}} & 0 & 0 & 0 & \frac{F}{3(2-F)}
\end{pmatrix}
\end{equation}
For any of the above state the extension can be found by means of
linear optimisation with help of SEDUMI module \cite{SEDUMI}. We
have found the extension of $\rho_{AB}$ very easily, in fact we
have for $F\leq\frac{1}{2}$ the following spectral decomposition
of the extension $\rho_{BAB}$:
\begin{equation}
   \left\lbrace
           \begin{array}{l l l}
          |\varphi_{0}\rangle=|020\rangle \textrm{ and } \lambda_{0}=\frac{1-F}{6}\\
|\varphi_{1}\rangle=|001\rangle+|100\rangle+|111\rangle+|122\rangle+|221\rangle
\textrm{ and }
\lambda_{1}=\frac{F}{3}\\
  |\varphi_{2}\rangle=|021\rangle \textrm{ and }
\lambda_{2}=\frac{1-2F}{6}\\
  |\varphi_{3}\rangle=|101\rangle \textrm{ and }
\lambda_{3}=\frac{1-2F}{3}\\
  |\varphi_{4}\rangle=|120\rangle \textrm{ and }
\lambda_{4}=\frac{1-F}{6}\\
  |\varphi_{5}\rangle=|121\rangle \textrm{ and }
\lambda_{5}=\frac{1-2F}{6}\\
          \end{array}
         \right.
\end{equation}
where generally eigenvalues have to fulfil following conditions so
that after tracing out Brigitte we obtain $\rho_{AB}$:

\begin{equation}
   \left\lbrace
           \begin{array}{l}
\lambda_{0} +
\lambda_{4}=\frac{1-F}{3}\\
\lambda_{2} +
\lambda_{5}=\frac{1-2F}{3}\\
                \end{array}
         \right.
\end{equation}

According to these constructions we may find another state
$\rho_{BAB}$ that is nearest (in the set of states constructed on
above eigenvectors) to singlet in sense of fidelity ($\cal
F=\langle\Psi_{+}|\rho_{AB}|\Psi_{+}\rangle$) of its local
reduction $\rho_{AB}$:
\begin{equation}
 \left\lbrace
           \begin{array}{l}
  \rho_{BAB}=\frac{1}{5}
  |\varphi_{1}\rangle\langle\varphi_{1}| \\
  \rho_{AB}=\frac{3}{5}P_{+} + \frac{1}{5}|01\rangle\langle01| +
  \frac{1}{5}|21\rangle\langle21|
   \end{array}
         \right.
\end{equation}

As a generalization of such states we construct states extreme in
the above sense for arbitrary dimension:
\begin{equation}
\Upsilon=\frac{d}{2d-1}P_{+}+\frac{1}{2d-1}\sum^{d-1}_{i=1}|i\;0\rangle\langle
i\;0|
\end{equation}
We state now the following question as a natural conclusion of
above analysis: \\{\textbf{Question:}} What is the maximal
possible value of fidelity of $\rho$ that we may obtain from
states for which $Q_{\rightarrow}=0$?

\section{Upper bound on \textit{$D_{\rightarrow}$}}

In this section we consider the distance of any state from the set
of extendible states. Note that the set of extendible states is
convex and compact, what can be obviously obtained from the
extendibility of any convex combination of extendible states.
Subsequently, we show that the set is closed under local
operations and one-way classical communication ($1$-LOCC) in the
following lemma:
\begin{lem}
The set $\cal E_{AB}$ of symmetrically extendible states is mapped
under $1$-LOCC for $\Lambda: B(\cal{H}_{AB})\rightarrow
B(\cal{H}_{\widetilde{A}\widetilde{B}})$ into the set of
symmetrically extendible states $\cal
E_{\widetilde{A}\widetilde{B}}$.
\end{lem}
\begin{proof}
 \begin{eqnarray*}
   \rho_{AB} \subset \cal E_{AB} \Rightarrow \exists_{\rho_{ABB'}} \; \rho_{ABB'}=\rho_{AB'B} \wedge Tr_{B'}\rho_{ABB'}=\rho_{AB}\\
   \Rightarrow Tr_{\widetilde{B'}}\Lambda(\rho_{ABB'})=\rho_{\widetilde{A}\widetilde{B}}\subset
\cal E_{\widetilde{A}\widetilde{B}}
  \end{eqnarray*}
where
\begin{eqnarray}
\Lambda(\rho_{ABB'})&=&\sum_{i,j=1}^{K,L}(I_{2}^{\widetilde{A}}\otimes
W_{ji}^{B\rightarrow\widetilde{B}} \otimes
W_{ji}^{B'\rightarrow\widetilde{B'}})(V_{i}^{A\rightarrow\widetilde{A}}\otimes
I_{1}^{B}\otimes
I_{1}^{B'})\rho_{ABB'}\times \\
& &\times (V_{i}^{A\rightarrow\widetilde{A}\dag}\otimes
I_{1}^{B}\otimes I_{1}^{B'})(I_{2}^{\widetilde{A}}\otimes
W_{ji}^{B\rightarrow\widetilde{B}\dag}\otimes
W_{ji}^{B'\rightarrow\widetilde{B'}\dag})\nonumber
\end{eqnarray}
and operations acting on Bob's side are trace-preserving due to
the necessity of non-breaking the property of extendibility.
\end{proof}

For our analysis we define the measure of this distance based on
the definition of relative entropy:
\begin{defn}
Assume that a convex set $\cal E_{AB}$ is a set of extendible
states, i.e.\begin{equation} \cal E_{AB}=\{\sigma_{AB}:
\exists_{\Psi_{ABB'C}}\;
 \sigma_{AB}=\sigma_{AB'}=Tr_{CB}[|\Psi_{ABB'C}\rangle\langle\Psi_{ABB'C}|] \}.
\end{equation}
Then the distance of a state $\rho_{AB}$ on
$\cal{H}_{AB}=\cal{H}_{A}\otimes \cal{H}_{B}$ with
$\dim\cal{H}_{A}=d_{A}$ and $\dim\cal{H}_{B}=d_{B}$ from the set
of extendible states $\cal E_{AB}$ of $d\otimes d$ type where
$d=\max[d_{A}, d_{B}]$ is defined by
\begin{equation}\label{extrelative}
R_{\cal E_{AB}}(\rho_{AB})=\delta_{AB}\inf_{\sigma_{AB}\in \cal E}
R(\widetilde{\rho}_{AB}\|\sigma_{AB})
\end{equation}
where $\forall_{\rho,
\sigma}\;R(\rho\|\sigma)=Tr[\rho\log\rho-\rho\log\sigma]$ and
$\delta=-\frac{\log d}{\log\frac{(d+1)}{2d}}$ with $d=\max[d_{A},
d_{B} ]$ due to normalization of this function on maximally
entangled states. In the formula (\ref{extrelative})
$\widetilde{\rho}_{AB}$ is taken as a state of $d\otimes d$ type
(after embedding $\rho_{AB}$ into $d\otimes d$ space).
\end{defn}

Using techniques \cite{Wer} we show that the nearest one in
arbitrary dimension is a state $\rho(d,F_{max})$ from subset of
isotropic states $\rho(d,F)$ \cite{MHPH99} with fidelity $F\leq
F_{max}$ for which those are symmetrically extendible:
\begin{equation}
  F_{max}=\frac{d+1}{2d}
\end{equation}
\begin{equation}
  \rho(d,F)=\frac{d^2}{d^2-1}[(1-F)\frac{I}{d^2}+(F-\frac{1}{d^2})P_{+}]
\end{equation}
Indeed, following \cite{Wer} one needs to analyze operators from a
six dimensional non-commutative $\mathbf{C}^{\ast}$-algebra that
are $\overline{U} \otimes U \otimes U$-invariant and
$V_{(23)}$-invariant. Such operators $S$ will be represented as a
linear combination of the basis elements of the algebra:
$B=\{S_{+}, S_{-}, S_{0}, S_{1}, S_{2}, S_{3}\}$ where for the
trace condition one obtains \cite{Wer} conditions for factors of
the combination: $s_{2}=s_{3}=0$ and, further, from positivity:
$s_{0}=1-s_{+}-s_{-}$.
\begin{equation}
 S=s_{+}S_{+}+s_{-}S_{-}+s_{0}S_{0}+s_{1}S_{1}
\end{equation}
The matter of interest is now the tetrahedron in three-dimensional
euclidian space of parameters $(s_{+}, s_{-}, s_{1})$ confined by
the hyperplanes \cite{Wer}: $\{h_{1}^{'}, h_{2}^{'}, h_{3}^{'},
h_{4}^{'}\}$ in which exists the state $\Omega_{ABE}$ giving the
searched symmetrically extendible reduction $\rho_{AB}$. For
maximizing the distance of the unknown state $\rho_{AB}$ to
singlet it suffices to find the maximization over fidelity
$\widetilde{F}$ between the symmetric extension represented as
$\Omega_{ABE}$ and virtually extended unnormalized operator
$\rho_{ABB'}=P_{+}\otimes I$ as
$\widetilde{F}_{max}=Tr[P_{+}\otimes I\; \Omega_{ABE}]=Tr[P_{+}\;
\rho_{AB} ]=F_{max}$ :
\begin{equation}
   \left\lbrace
           \begin{array}{l}
F_{+}=Tr[(P_{+}\otimes I)\; S_{+}]/Tr[S_{+}^2]=0 \\
F_{-}=Tr[(P_{+}\otimes I)\; S_{-}]=0 \\
F_{0}=Tr[(P_{+}\otimes I)\; S_{0}]/Tr[S_{0}^2]=d/2d \\
F_{1}=Tr[(P_{+}\otimes I)\; S_{1}]/Tr[S_{1}^2]=1/2d \\
     \end{array}
         \right.
\end{equation}
\begin{equation}
   \left\lbrace
           \begin{array}{l}
  \widetilde{F}=F_{0}+\overrightarrow{s}\circ\overrightarrow{f} \\
  \widetilde{F}_{max}=\max_{\overrightarrow{s}\in\Delta}\widetilde{F} \\
\end{array}
         \right.
\end{equation}
where $\Delta$ denotes the tetrahedron bounded by mentioned
hyperplanes, $\overrightarrow{f}=[F_{+}-F_{0}, F_{-}-F_{0},
F_{1}]$ and $\overrightarrow{s}=[s_{+}, s_{-}, s_{0}]$.
Normalization of parameters $F_{i}$ inherits from the commutation
relations \cite{Wer} between operators $S_{i}$. Maximization
results in $\overrightarrow{s}=[0, 0, 1]$ that relates to the
found aforementioned isotropic states $\rho_{AB}=\rho(d,F_{max})$.
The explicit form of the tripartite symmetric extension of
isotropic states $\rho(d,F_{max})$ in the border of extendibility
is:
\begin{equation}
  \Omega_{ABE}=\frac{1}{2d}(S_{0}+S_{1})
\end{equation}
where \cite{Wer}:
\[
S_{0}=\frac{1}{d^2-1}(d(X+VXV)-(XV+VX)) \;\;and\;\;
S_{1}=\frac{1}{d^2-1}(d(XV+VX)-(X+VXV))\] for
\[|\Phi\rangle=\sum_{i}|ii\rangle \; , \;
X=|\Phi\rangle\langle \Phi|\otimes I \; , \;
V=V_{(23)}=\sum_{ijk}|ijk\rangle\langle ikj|.
\]

It is important to notice that the same results can be obtained
numerically by means of linear programming methods that we have
utilized to find the broad class of symmetrically extendible
states.

Following we analyze if similarly to distance from separable
states one can construct an appropriate entanglement measure
basing on (\ref{extrelative}). The normalized distance from the
set of extendible states does not satisfy though all necessary
conditions \cite{PLE97,PLE98} that every measure of one-way
distillable entanglement has to satisfy: introduction of the
normalization factor $\delta_{AB}$ causes that $R_{\cal
E_{AB}}(\rho)$ becomes explicitly dependant on the dimension of
the system $AB$, therefore, for protocols increasing dimension of
input state the parameter is not a monotone:
\begin{description}
  \item[A1.]If $\sigma_{AB}$ is separable then
  $R_{\cal E_{AB}}(\sigma_{AB})=0$ due to the fact that every
  separable state is extendible.
  \item[A2.]Local unitary operations leave
  $R_{\cal E_{AB}}(\sigma_{AB})$ invariant, that is satisfied due to
  invariancy of distance measures under local unitary
  transformations, i.e. $R_{\cal E_{AB}}(\sigma_{AB})=R_{\cal E_{AB}}(U_{A}\otimes U_{B}\sigma_{AB}U_{A}^{\dag}\otimes
  U_{B}^{\dag})$.
  \item[A3.](Restricted $1$-LOCC monotonicity.) The parameter $R_{\cal E_{AB}}(\sigma_{AB})$ of one-way
  distillable entanglement does not increase under non-increasing dimension $1$-LOCC, i.e.
  $\Lambda: B(\cal{H}_{AB})\rightarrow
B(\cal{H}_{\widetilde{A}\widetilde{B}})$ with $n_{AB}=\max[d_{A},
d_{B} ]$, $n_{\widetilde{A}\widetilde{B}}=\max[d_{\widetilde{A}},
d_{\widetilde{B}} ]$ for $n_{AB}\geq
n_{\widetilde{A}\widetilde{B}}$, then
\begin{equation}
  R_{\cal E_{\widetilde{A}\widetilde{B}}}(\Lambda\sigma_{AB})\leq R_{\cal E_{AB}}(\sigma_{AB})
\end{equation}
    This condition may be simply proved due to non-increasing of
    $R(\rho\|\sigma)$ under a subclass of $1$-LOCC operations $\Lambda$ that is stated
    above in the lemma. Namely, because $\Lambda \cal E_{AB} \subset
    \cal E_{\widetilde{A}\widetilde{B}}$ and assuming that $\sigma^{*}$ is an extendible state
    that realizes the minimal value in eq.(\ref{extrelative}) we have:
\begin{equation}
  R_{\cal E_{AB}}(\rho)=\delta_{AB} R(\rho\|\sigma^{*})\geq
  \delta_{\widetilde{A}\widetilde{B}} R(\Lambda\rho\|\Lambda\sigma^{*})\geq \delta_{\widetilde{A}\widetilde{B}} \inf_{\sigma \in \cal E_{\widetilde{A}\widetilde{B}}}
R(\Lambda \rho\|\sigma)=R_{\cal
E_{\widetilde{A}\widetilde{B}}}(\Lambda\rho)
\end{equation}
where $n_{AB}\geq n_{\widetilde{A}\widetilde{B}}$ derives the
condition $\delta_{AB}\geq \delta_{\widetilde{A}\widetilde{B}}$.
\end{description}
However, we show further that the entanglement parameter can be
utilized for bounding one-way entanglement of distillation due to
preparation of the measure in asymptotic regime.

In general, every entanglement parameter of type $E(\sigma)=
\alpha \inf_{\rho \in \Delta} \cal D(\sigma\|\rho) $ where $\cal
D(\sigma\|\rho)$ is appropriate distance between $\sigma$ and
$\rho$, $\Delta$ denotes the characteristic set to which the
distance is measured and $\alpha$ normalizes the parameter so that
$E(|\Psi_{+}\rangle\langle \Psi_{+}|)=\log d$ is not monotonic,
i.e. $\exists_{\Lambda} \; E(\sigma) > E(\Lambda(\sigma))$. For
$R_{\cal E_{AB}}$ unitary injection of input state $\rho_{AB}$
into higher dimensional space gives $R_{\cal E_{AB}}(\rho) >
R_{\cal E_{\widetilde{A}\widetilde{B}}}(\Lambda(\rho))$.

Additionally, following analysis in \cite{MH01,DO01}, we show that
the entanglement parameter satisfies:
\begin{description}
\item[B1.](Continuity on isotropic states.)
We may simply show that this parameter is continuous on isotropic
states $\rho(d_{n}, F_{n})$ with $F_{n}\rightarrow 1,
d_{n}\rightarrow \infty$ that means
\[
\frac{R_{\cal E}(\rho(d_{n}, F_{n}))}{\log d_{n}}\rightarrow 1
\]
as then $R_{\cal E}(\rho(d_{n}, F_{n}))\rightarrow \log d_{n}$
that is easy to check.

\end{description}

Following the papers \cite{MH01,PRMH00} and the above definition
we define the distance in asymptotic regime as follows:
\begin{equation}\label{regularization}
R_{\cal
E_{AB}}^{\infty}(\rho_{AB})=\limsup_{n\rightarrow\infty}\frac{R_{\cal
E_{AB}}({\rho_{AB}}^{\otimes n})}{n}
\end{equation}

One can also propose other measures, but this will be subject of
analysis elsewhere \cite{mesaures}.

Having defined above regularized parameter $R_{\cal
E_{AB}}^{\infty}(\rho_{AB})$, we are able now to determine the
upper bound on the one-way distillable entanglement. In
\cite{DV04} Devetak and Winter have proved very powerful
conjecture called "hashing inequality"\[ D_{\rightarrow} \geq
S(\rho_{B})-S(\rho_{AB})
\] from which one may find particular states of non-zero
\textit{$D_{\rightarrow}$}. For the very features of measures that
bound the distillable entanglement \textit{$D_{\rightarrow}$},
defined in \cite{MH01,DO01}, where was shown that monotonicity,
continuity on isotropic states are sufficient for any properly
regularised function to be upper bound for $D_{\rightarrow}$, we
may prove now the following theorem exploiting only distillation
protocols in the line of the proof:

\begin{thm}
For any bipartite state $\rho_{AB}$ there holds:
\begin{equation}
D_{\rightarrow}(\rho_{AB})\leq R_{\cal E_{AB}}^{\infty}(\rho_{AB})
\end{equation}
\end{thm}

\begin{proof}
Any one-way distillation protocol can be reduced to the
distillation protocol \cite{MH01,DO01,PRMH00} where the input is
$\rho^{\otimes n}$ and the output is a family of the states
$\rho(d_{n}, F_{n})$ with $\lim_{n\rightarrow \infty}\frac{\log
d_{n}}{n}=D_{\rightarrow}(\rho)$ and $F_{n}\rightarrow 1$. We may
always put $d_{n}\leq n_{AB}^{n}$ for $n_{AB}=\min[d_{A}, d_{B}]$
since there holds $D_{\rightarrow}(\rho)\leq \log n_{AB}$. Thus,
we can consider only $1$-LOCC non-increasing dimensions of input
and so monotonicity of $R_{\cal E_{AB}}$ holds. By analogy with
the theorem put in \cite{MH01,DO01,PRMH00} the properties (A$3$)
and (B$1$) imply that $R_{\cal E_{AB}}^{\infty}(\rho_{AB})$ is
upper bound for $D_{\rightarrow}$. The regularisation
(\ref{regularization}) with supreme value enables upper bound of
$D_{\rightarrow}$.
\end{proof}

\section{Conclusions}

Quantum channels theory still has many unsolved problems. We have
pointed out a general test for zero capacity of one-way  channel
capacity (which has been shown to be equal to zero-way capacity
\cite{Barnum}). The test is based on checking of the existence of
symmetric extension of a state isomorphic to a given channel. The
test can be very easily performed with help of popular
semi-definite programming codes. Finally, basing on the test, we
have found a new parameter of entanglement. Its suitable
regularized version is an upper bound on one-way distillable
entanglement of given quantum state. Note that, although the
entanglement monogamy property was known for a long time, this is
the first entanglement parameter basing explicitly on that
property and the symmetric extension of quantum states. We hope
that the above results will help in further analysis of various
aspects of quantum channels. It is very interesting that recently
developed complete hierarchies approach to separability problem
\cite{JE} has been extended \cite{PC} to include symmetric
extensions of quantum operators, which leads to class of
entanglement measures. This gives a hope that symmetric extensions
will be a useful tool not only to qualify but also to quantify
some aspects of quantum entanglement.


\begin{thebibliography}{20}
\bibitem{huge}
C. H. Bennett, D. P. DiVincenzo, J. A. Smolin and W. K. Wootters
Phys. Rev. A {\bf 5}, 3824 (1996)
\bibitem{Springer}
H.-K. Lo, S. Popescu and T. Spiller, (eds.) {\it Introduction in
quantum information and computation}, World Scientific, (1998); J.
Gruska, {\it Quantum Computing}, McGraw-Hill, London 1999; D.
Bouwmeester, A. K. Ekert, A. Zeilinger (eds.), {\it The physics of
quantum information : quantum cryptography, quantum teleportation,
quantum computation}; Springer, New York 2000; M. A. Nielsen and
I. L. Chuang, {\it Quantum Computation and Quantum Information},
Cambridge University Press, Cambridge 2000; G.~Alber, T.~Beth,
M.~Horodecki, P.~Horodecki, R.~Horodecki, M.~R\"ottler,
H.~Weinfurter, R.~F. Werner, and A.~Zeilinger, {\em Quantum
information: an introduction to basic theoretical concepts and
experiments.}, volume 173 of {\em Springer Tracts in Modern
Physics}, Springer, Berlin 2001.

\bibitem{distillation}
 C. H. Bennett, G. Brassard, S. Popescu, B. Schumacher, J. A.
Smolin and W. K. Wootters, Phys.Rev.Lett. {\bf 76}, 722 (1996).
\bibitem{Bruss}
 D. Bruss, D. P. DiVincenzo, A. Ekert, C. A. Fuchs, C. Macchiavello and J. A. Smolin,
Phys. Rev. A, {\bf 57}, 2368 (1998).
\bibitem{Barnum0}
H. Barnum, E. Knill and M. A. Nielsen, IEEE Trans.Info.Theor. {\bf
46} 1317 (2000).
\bibitem{channels}
M. Horodecki, P. Horodecki and R.Horodecki, Phys. Rev. Lett. {\bf
85}, 433 (2000).
\bibitem{Shor} P. W. Shor,
Capacities of quantum channels and how to find them, Preprint
quant-ph/0304102.
\bibitem{Devetak}I. Devetak I, The private
classical information capacity and quantum information capacity of
a quantum channel, Preprint quant-ph/0304127.
\bibitem{DV04}
I. Devetak and A. Winter, Phys. Rev. Lett. {\bf 93}, 080501
(2004).
\bibitem{PH03}
P. Horodecki, Centr. Eur. J. Phys. {\bf 4}, 695 (2003).
\bibitem{DCH04}
W. D\"ur, J. I. Cirac and P. Horodecki, Phys. Rev. Lett. {\bf 93},
020503 (2004).

\bibitem{Cerf}
N. J. Cerf, Phys. Rev. Lett. {\bf 84}, 4497 (2000).

\bibitem{D1}
A. C. Doherty, P. A. Parillo and F. M. Spedalieri, Phys. Rev.
Lett. {\bf 88}, 187904 (2002) .
\bibitem{D2}
A. C. Doherty, P. A. Parillo and F. M. Spedalieri, Phys. Rev. A
{\bf 69}, 022308 (2004).
\bibitem{T1}
B. M. Terhal, A. C. Doherty and D. Schwab, Phys. Rev. Lett. {\bf
90}, (2003).

\bibitem{Jam72}
A. Jamiolkowski, Rep. Math. Phys. {\bf 3}, 275 (1972).
\bibitem{Choi75}
M. D. Choi, Linear Algebra and Its Applications {\bf 10}, 285
(1975).

\bibitem{Barnum}
H. Barnum, J. A. Smolin and B. M. Terhal, Phys. Rev. A {\bf 58},
3496 (1998).
\bibitem{DiV}
C. H. Bennett, D. P. DiVincenzo, and J. A. Smolin, Phys. Rev.
Lett. {\bf 78}, 3217 (1997).
\bibitem{Vidal}
G. Vidal, On the continuity of asymptotic measures of
entanglement, Preprint quant-ph/0203107.


\bibitem{BMH04}
G. Brassard, P. Horodecki and T. Mor, IBM J. Res. Dev. {\bf 84},
87 (2004).


\bibitem{tcv}
D. Kretschmann and R. F. Werner, New J. Phys. {\bf 6} 26 (2004).
\bibitem{SEDUMI}
J. Sturm, SEDUMI VERSION 1.05.2001,
http://fewcal.kub.nl/sturm/software/sedumi.html.
\bibitem{PLE97}
V. Vedral, M. B. Plenio, M. A. Rippin and P. L. Knight, Phys. Rev.
Lett. {\bf 78}, 2275 (1997).
\bibitem{PLE98}
V. Vedral and M. B. Plenio, Phys. Rev. A {\bf 57}, 1619 (1998).
\bibitem{Wer}
T. Eggeling, R. F. Werner, Phys. Rev. A {\bf 63}, 042111 (2001).
\bibitem{MHPH99}
M. Horodecki and P. Horodecki, Phys. Rev. A {\bf 59}, 4206 (1999).
\bibitem{MH01}
M. Horodecki, Quant. Info. Comp. 1, 1 (2001).
\bibitem{DO01}
M. J. Donald, M. Horodecki and O. Rudolph, J. Math. Phys. {\bf
43}, 4252 (2002).
\bibitem{PRMH00}
M. Horodecki, P. Horodecki and R.Horodecki, Phys. Rev. Lett. {\bf
84}, 2014 (2000).
\bibitem{mesaures}
For instance one can propose fidelity of state according to the
nearest purified extension as follows $F_{\cal
E}(\rho_{AB})=\inf_{\sigma_{AB}\in \cal
E}F(\rho_{AB},\sigma_{AB})=\inf_{\sigma_{AB}\in \cal E
}|\langle\Phi_{ABB'C}|\Psi_{ABB'C}\rangle|$ where the set $\cal E$
is defined as in above definition and both $\Phi_{ABB'C}$ and
$\Psi_{ABB'C}$ is a purification of a suitable state. It can be
shown that such quantity is also restricted $1$-LOCC monotone.
\bibitem{DH99}
M. J. Donald and M. Horodecki, Phys. Lett. A {\bf 264}, 257
(1999).
\bibitem{PC}
J. Eisert, private communication.
\bibitem{JE}
J. Eisert, P. Hyllus, O. Guehne and M. Curty, Phys. Rev. A {\bf
70}, 062317 (2004).

\end{thebibliography}
\end{document}